\documentclass[%
 reprint,
 amsmath,amssymb,
 aps,
]{revtex4-2}

\usepackage{graphicx}
\usepackage{dcolumn}
\usepackage{bm}

\usepackage[usenames]{color}

\begin{document}

\preprint{APS/123-QED}

\title{Torsion-induced Dzyaloshinskii-Moriya interaction in helical magnets}

\author{M.A. Kuznetsov}
 \email{kuznetsovm@ipmras.ru}

 \author{A.A. Fraerman}%
 \email{andr@ipmras.ru}
\affiliation{%
 Institute for Physics of Microstructures, Russian Academy of Sciences, Akademicheskaya St. 7,
Nizhny Novgorod 607680, Russian Federation
}%


\begin{abstract}
It has been shown that in magnets possessing an inversion center in the absence of deformations, a torsion-induced Dzyaloshinsky-Moriya interaction (tiDMI) can arise. A microscopic mechanism for this interaction is described, involving the transfer of angular momentum to the lattice upon electron reflection from the magnet's boundary. An estimate of the tiDMI constant is provided. It is demonstrated that tiDMI can lift the chiral degeneracy in helimagnets, and a way for experimentally observing this effect is proposed. 
\end{abstract}

\maketitle


\section{Introduction}

Magneto-mechanical phenomena, first investigated in the pioneering works of Einstein--de Haas~\cite{einstein1915} and Barnett~\cite{barnett1915}, continue to attract researchers’ attention due to their potential for developing novel micro- and nanomechanical systems. For instance, mechanical deformations induced by spin-polarized electric currents in nanowires have been studied  theoretically in works~\cite{fulde1998,mohanty2004,malsukov2005,yu2007}, and the mechanisms of mechanical generation of spin current have also been described~\cite{matsuo2015}. Deformations can arise not only under electric current flow but also under equilibrium conditions if an equilibrium spin current exists in the system. Such a situation is possible, for example, in a conducting medium with Rashba spin--orbit coupling~\cite{sonin2007}. In fact, that work proposed a method for visualizing equilibrium spin currents based on the bending of a cantilever caused by the variation of the spin current---from zero at the free edge of the cantilever to a constant value far from this edge. This approach assumes that the change in electron spins due to spin--orbit interaction is transferred to the crystal lattice, and the resulting mechanical torque induces cantilever bending.

In the present work, we consider another system that supports an equilibrium spin current—namely, a system with a noncollinear  magnetization distribution $\mathbf{M}$, described by the unit vector $\mathbf{n} = \mathbf{M}/M_s$, where $M_s$ is the saturation magnetization. It is known~\cite{sonin2010} that the phenomenological expression for the spin-current density tensor in such a system is given by
\begin{equation}
\mathcal{Q}_{i\alpha} = \left( \mathbf{n} \times \frac{\partial \mathbf{n}}{\partial x_i} \right)_\alpha, \label{eq:Q_def}
\end{equation}
where an arbitrary prefactor has been set to unity. The components of this tensor are proportional to the flux of electrons carrying spin component $\alpha$ and velocity component $i$. Using Eq.~\eqref{eq:Q_def}, one can readily compute the spin-current density tensor for a helical magnetization profile $\mathbf{n} = (\cos qz, \sin qz, 0)$, with $q = 2\pi/\lambda$ and $\lambda$ being the helix period. In this case, the only nonvanishing component is $\mathcal{Q}_{zz} = q$.

The tensor $\mathcal{Q}_{i\alpha}$ can contribute to the magnetic energy density if the crystal symmetry allows the existence of a second-rank pseudotensor $T_{i\alpha}$, which may possess both symmetric and antisymmetric parts. In an elastically isotropic medium, whose deformations are described by the displacement vector $\mathbf{u}$, such a pseudotensor can be constructed as follows:
\begin{align}
T_{i\alpha}&=\frac{1}{2}D_1\left( \frac{\partial}{\partial x_i} (\nabla\times\mathbf{u})_\alpha + \frac{\partial}{\partial x_\alpha} (\nabla\times\mathbf{u})_i \right)\nonumber\\ 
&+\frac{1}{2} D'_2 \, \epsilon_{i\alpha n} \frac{\partial}{\partial x_n} (\nabla\cdot\mathbf{u}) \nonumber\\
&
+\frac{1}{2} D'_3 \, \epsilon_{i\alpha n} \left(\nabla\times(\nabla\times\mathbf{u})\right)_n, \label{eq:T}
\end{align}
where $\epsilon_{i\alpha n}$ is the Levi-Civita symbol, and $D_1$, $D'_2$, $D'_3$ are constants. Note that expression~\eqref{eq:T} is derived assuming small deformations that vary slowly in space~\cite{landau1986}. Each of the three terms in Eq.~\eqref{eq:T} is independent and cannot be expressed through the other two. However, in the absence of body forces, the equilibrium equation for an elastically isotropic medium holds~\cite{landau1986}:
\begin{equation}
\nabla\times(\nabla\times\mathbf{u}) = \frac{2(1 - \nu)}{1 - 2\nu} \, \nabla\,(\nabla\cdot\mathbf{u}), \label{eq:equilibrium}
\end{equation}
where $\nu$ is Poisson’s ratio. This relation allows us to connect the last two terms in Eq.~\eqref{eq:T}. Consequently, the energy density of the deformation-induced Dzyaloshinskii--Moriya interaction (diDMI), $\varepsilon_{\text{diDMI}}$, can be written as:
\begin{align}
\varepsilon_{\text{diDMI}}&=\frac{1}{2} D_1 \left( \frac{\partial}{\partial x_i} (\nabla\times\mathbf{u})_\alpha + \frac{\partial}{\partial x_\alpha} (\nabla\times\mathbf{u})_i \right) \mathcal{Q}_{i\alpha}\nonumber \\
&+\frac{1}{2} D_2 \, \epsilon_{i\alpha n} \left( \frac{\partial}{\partial x_n} (\nabla\cdot\mathbf{u}) \right) \mathcal{Q}_{i\alpha}, \label{eq:ediDMI}
\end{align}
where $D_2 = D'_2 + 2D'_3 (1 - \nu)/(1 - 2\nu)$. Thus, inhomogeneous deformations can break chiral magnetic symmetry and induce Dzyaloshinskii--Moriya interaction~\cite{dzyaloshinskii1957,moriya1960} in a magnet that, in the absence of deformations, possesses an inversion center. The possibility of chiral symmetry breaking due to inhomogeneous deformations was previously noted in works~\cite{fedorov1997,kitchaev2018,fraerman2023}, and experimental evidence supporting this hypothesis has been reported in~\cite{zhang2021,liu2022}. However, the physical meaning of the terms in Eq.~\eqref{eq:ediDMI} has not been discussed, and estimates for the constants $D_1$ and $D_2$ governing the strength of diDMI are still lacking.

The second term in Eq.~\eqref{eq:ediDMI} describes chiral symmetry breaking caused by nonuniform compression or extension---for example, arising under bending deformations~\cite{landau1986}. In the present work, however, we focus exclusively on pure torsional deformations and the associated torsion-induced Dzyaloshinskii--Moriya interaction (tiDMI). For a helical magnetization distribution, the second (``antisymmetric'') term in Eq.~\eqref{eq:ediDMI} vanishes, and the tiDMI energy density takes the form:
\begin{align}
\varepsilon_{\text{tiDMI}}&=D_1 \left( \frac{\partial}{\partial z} (\nabla\times\mathbf{u})_z \right) \mathcal{Q}_{zz} \nonumber\\
&= D_1 \, q \, \frac{\partial}{\partial z} (\nabla\times\mathbf{u})_z. \label{eq:etiDMI}
\end{align}
We thus observe a direct coupling between spin currents in the helix and torsional deformations about its axis.

In this paper, we investigate the magneto-mechanical coupling in chiral magnetic systems arising from the interaction between pure torsional deformations and the spin current, which acts as a torque on the crystal lattice of the magnet. Within the framework of the $s$--$d$ exchange model, we calculate the magnitude of this effect and provide estimates for the constant $D_1$ that determines the strength of tiDMI.

\section{Microscopic mechanism of torsion-induced Dzyaloshinskii--Moriya interaction}

Consider a magnetic rod of length $L$ with a helical magnetization distribution $\mathbf{n} = (\cos qz, \sin qz, 0)$. The $z$-axis is aligned along the rod axis, which coincides with the axis of the magnetic helix. We assume that the rod length $L$ greatly exceeds the helix period $\lambda$, i.e., $L \gg \lambda$. One end of the rod ($z = L$) is rigidly fixed, while the other ($z = 0$) is free (Fig.~1a). Within the $s$--$d$ model, electron dynamics in the magnetic helix is governed by the Schr\"odinger equation with the Hamiltonian (see, for example, a brief overview~\cite{Irkhin2022})
\begin{equation}
\hat{H} = -\frac{\hbar^2}{2m} \Delta + J\, (\hat{\boldsymbol{\sigma}} \cdot \mathbf{n}), \label{eq:H}
\end{equation}
where $m$ is the electron effective mass, $J$ is the $s$--$d$ exchange constant, and $\hat{\boldsymbol{\sigma}}$ is the vector of Pauli matrices. From the time-dependent Schr\"odinger equation with Hamiltonian~\eqref{eq:H}, one obtains the equation for the spin density $\mathbf{s} = (\hbar/2)\, \psi^\dagger \hat{\boldsymbol{\sigma}} \psi$ (see, e.g.,~\cite{sonin2007}):
\begin{subequations}
\begin{align}
\frac{\partial s_\alpha}{\partial t} &+ \frac{\partial Q_{i\alpha}}{\partial x_i} = G_\alpha, \label{eq:spin_cont_a} 
\\
Q_{i\alpha} &= -\frac{i\hbar^2}{4m} \left( \psi^\dagger \hat{\sigma}_\alpha \frac{\partial \psi}{\partial x_i} - \frac{\partial \psi^\dagger}{\partial x_i} \hat{\sigma}_\alpha \psi \right), \label{eq:spin_cont_b}
\\
G_\alpha &= \frac{2J}{\hbar} \left( \mathbf{n} \times \mathbf{s} \right)_\alpha, \label{eq:spin_cont_c}
\end{align}
\end{subequations}
where $\psi$ is the two-component wave function (spinor) of the electron, and $Q_{i\alpha}$ is the spin-current density tensor. In the stationary state, $\partial s_\alpha / \partial t = 0$, but the spin is generally not conserved; the spatial inhomogeneity of the spin current is compensated by the ``torque'' $G_\alpha$. Accounting for the conservation of the total angular momentum—the sum of the electron spins and the lattice angular momentum—one may assume that the lattice experiences a torque $\mathbf{T}$ equal in magnitude and opposite in sign to the torque acting on the electrons, i.e., $T_\alpha = -G_\alpha$~\cite{sonin2007}. We shall show that $T_\alpha$ is nonzero near the free end of the helimagnetic rod and induces torsional deformation.

To do this, we first consider a region of the rod sufficiently far from its ends, where the rod can be treated as infinite. In an infinite sample, the Hamiltonian~\eqref{eq:H} is invariant under an arbitrary translation along the helix axis combined with a simultaneous spinor rotation about the same axis. This symmetry allows one to find the eigenfunctions and eigenvalues of the Hamiltonian~\cite{calvo1979}:
\begin{subequations}
\begin{align}
\textbf{\label{eq:psi}}
\psi_\pm &= \frac{1}{\sqrt{V(1 + \delta_\pm^2)}} 
\begin{pmatrix}
\delta_\pm e^{-i q z / 2} \\
e^{i q z / 2}
\end{pmatrix}
e^{i \mathbf{k} \cdot \mathbf{r}},\\ 
\label{eq:eps}
\varepsilon_\pm &= \frac{\hbar^2}{2m} \left( k^2 + \frac{q^2}{4} \pm k_J^2 \sqrt{ \left( \frac{q k_z}{k_J^2} \right)^2 + 1 } \right), 
\end{align}
\end{subequations}
where we have introduced
\begin{equation}
\delta_\pm = \operatorname{sgn}(J) \left( -\frac{q k_z}{k_J^2} \pm \sqrt{ \left( \frac{q k_z}{k_J^2} \right)^2 + 1 } \right). \label{eq:delta}
\end{equation}
Here, $V$ is the sample volume, $\mathbf{k}$ is the electron wave vector, $k_z$ is its $z$-component, and $k_J^2 = 2m |J| / \hbar^2$. The indices ``$+$'' and ``$-$'' correspond to minority (``$+$'') and majority (``$-$'') electrons, which possess different helicities (projections of spin onto momentum). For a helical magnetization distribution, only one diagonal component of the spin-current density tensor is nonzero: $Q_{zz}$. For majority and minority electrons, this component can be written as
\begin{equation}
Q_{zz}^\pm = \frac{\hbar}{m} \left( k_z s_z^\pm - \frac{\hbar}{2V} \frac{q}{2} \right), \label{eq:Qzz_pm}
\end{equation}
where the $z$-component of the spin density $s_z^\pm$ is
\begin{equation}
s_z^\pm = \mp \frac{\hbar}{2V} \frac{q k_z}{k_J^2} \frac{1}{\sqrt{ (q k_z / k_J^2)^2 + 1 }}. \label{eq:s_z_pm}
\end{equation}
It is straightforward to verify that all other components of the spin current and spin density vanish after summation over all allowed electron states. Thus, electron motion along the helix axis generates a $z$-component of spin. This component has opposite signs for the two spin subbands ($\varepsilon_+$ and $\varepsilon_-$) and depends on the $z$-component of the wave vector.

We now compute the total spin-current density produced by both minority and majority electrons:
\begin{equation}
\langle Q_{zz} \rangle_{\mathbf{k}} = \langle Q_{zz}^+ \rangle_{\mathbf{k}} + \langle Q_{zz}^- \rangle_{\mathbf{k}}, \label{eq:Q_total}
\end{equation}
where the brackets $\langle...\rangle_{\mathbf{k}}$ denote summation over all electron states with allowed wave vectors $\mathbf{k}$, i.e.,
\begin{equation}
\langle Q_{zz}^\pm \rangle_{\mathbf{k}} = \frac{V}{4\pi^2} \int\limits_{-k_z^{F\pm}}^{k_z^{F\pm}} \left( \int\limits_0^{k_\perp^{F\pm}(k_z)} k_\perp\, Q_{zz}^\pm\, dk_\perp \right) dk_z, \label{eq:Q_avg}
\end{equation}
with $k_\perp = (k_x^2 + k_y^2)^{1/2}$, and $k_\perp^{F\pm}$, $k_z^{F\pm}$ being the transverse and longitudinal Fermi wave vectors for the upper ($\varepsilon_+$) and lower ($\varepsilon_-$) spin subbands:
\begin{subequations}
\begin{align}
k_\perp^{F\pm}(k_z) &= \sqrt{ k_F^2 - \frac{q^2}{4} - k_z^2 \mp k_J^2 \sqrt{ 1 + \left( \frac{q k_z}{k_J^2} \right)^2 } }, \label{eq:k_perp_F} \\
k_z^{F\pm} &= \sqrt{ k_F^2 + \frac{q^2}{4} \mp \sqrt{ k_F^2 q^2 + k_J^4 } }, \label{eq:k_z_F}
\end{align}
\end{subequations}
where $k_F = \sqrt{2m \varepsilon_F / \hbar^2}$ and $\varepsilon_F$ is the Fermi energy. For the evaluation of integrals~\eqref{eq:Q_avg}, it is convenient to introduce the dimensionless parameters $\kappa = q / k_F$ and $\chi = |J| / \varepsilon_F$. In the limit $\kappa \ll 1$ and $\chi \ll 1$, the Fermi surface defined by Eqs.~\eqref{eq:k_perp_F} and~\eqref{eq:k_z_F} approaches a sphere of radius $k_F$. In this regime, the ratio $\kappa / \chi = q k_F / k_J^2$ determines the magnitude of the $z$-component of the spin density~\eqref{eq:s_z_pm} for electrons near the Fermi energy. Summing over all states under the conditions $\kappa \ll 1$ and $\chi \ll 1$, we obtain
\begin{equation}
\langle Q_{zz} \rangle_{\mathbf{k}} \approx -\frac{\hbar^2 N q}{16 m} \left( \frac{k_J}{k_F} \right)^4 = -\frac{\hbar^2 N q}{16 m} \left( \frac{J}{\varepsilon_F} \right)^2, \label{eq:Q_final}
\end{equation}
where $N = k_F^3 / (3\pi^2)$ is the electron concentration. Equation~\eqref{eq:Q_final} remains valid for arbitrary $\kappa / \chi$ provided that $k_F^2 > k_J^2 + q^2 / 4$ (or $\chi < 1 - \kappa^2 / 4$), ensuring that $\varepsilon_F$ lies above the bottom of the upper subband. Note that in an infinite rod, the torque $G_\alpha$ vanishes after summation over all allowed electron states.

In a finite sample, the combined translational-rotational symmetry along the helix axis is broken. Upon reflection from the boundary, electrons undergo spin-dependent scattering, which leads to non-collinearity of the transverse component of the spin density with respect to the magnetization $\mathbf{n}$ and, as a consequence, to the appearance of a non-zero $z$-component of the torque, $\langle G_z \rangle_{\mathbf{k}}$. Indeed, for a stationary state in a finite rod, Eq.~\eqref{eq:spin_cont_a} implies
\begin{equation}
\left. \langle Q_{zz} \rangle_{\mathbf{k}} \right|_{z=l} - \left. \langle Q_{zz} \rangle_{\mathbf{k}} \right|_{z=0} = \int\limits_0^l \langle G_z \rangle_{\mathbf{k}}\, dz, \label{eq:boundary_balance}
\end{equation}
where, for definiteness, the free end is located at $z = 0$ (Fig.~1a). Here, $l$ is a distance from the boundary satisfying $l \gg l_{\text{so}}$, where $l_{\text{so}}$ is the spin--orbit relaxation length over which angular momentum is transferred from electrons to the lattice, and simultaneously $l \ll L$, so that $L \gg l_{\text{so}}$. Thus, beyond distance $l$, the boundary influence is negligible, and $\langle Q_{zz} \rangle_{\mathbf{k}}|_{z=l}$ coincides with the spin-current density in the infinite sample (Eq.~\eqref{eq:Q_final}). Assuming an impenetrable barrier at the rod end, we set $\langle Q_{zz} \rangle_{\mathbf{k}}|_{z=0} = 0$. The macroscopic torque acting on the rod end,
\begin{equation}
\langle T_z \rangle = S \int\limits_0^l \langle T_z \rangle_{\mathbf{k}}\, dz=S \int\limits_0^{l_{\text{so}}} \langle T_z \rangle_{\mathbf{k}}\, dz,
\end{equation}
is then related to the total spin current in the infinite sample,
\begin{equation}
\langle Q_{zz} \rangle = S \int\limits_0^L \langle Q_{zz} \rangle_{\mathbf{k}}\, dz,
\end{equation}
by the relation
\begin{equation}
\langle T_z \rangle = -\frac{1}{L} \langle Q_{zz} \rangle, \label{eq:T_vs_Q}
\end{equation}
where $S$ is the rod cross-sectional area. Because $L \gg l_{\text{so}}$, the torque $\langle T_z \rangle$ can be considered applied directly at the rod boundary. Since this torque is directed along the rod axis, it induces torsional deformation (Fig.~1b). Notably, reversing the chirality (sign of $q$) of the magnetic helix (Fig.~1c) reverses the sign of the spin current~\eqref{eq:Q_final} and, consequently, the direction of torsion (Fig.~1d).

\begin{figure*}[t]
\includegraphics[width = 0.85\textwidth]{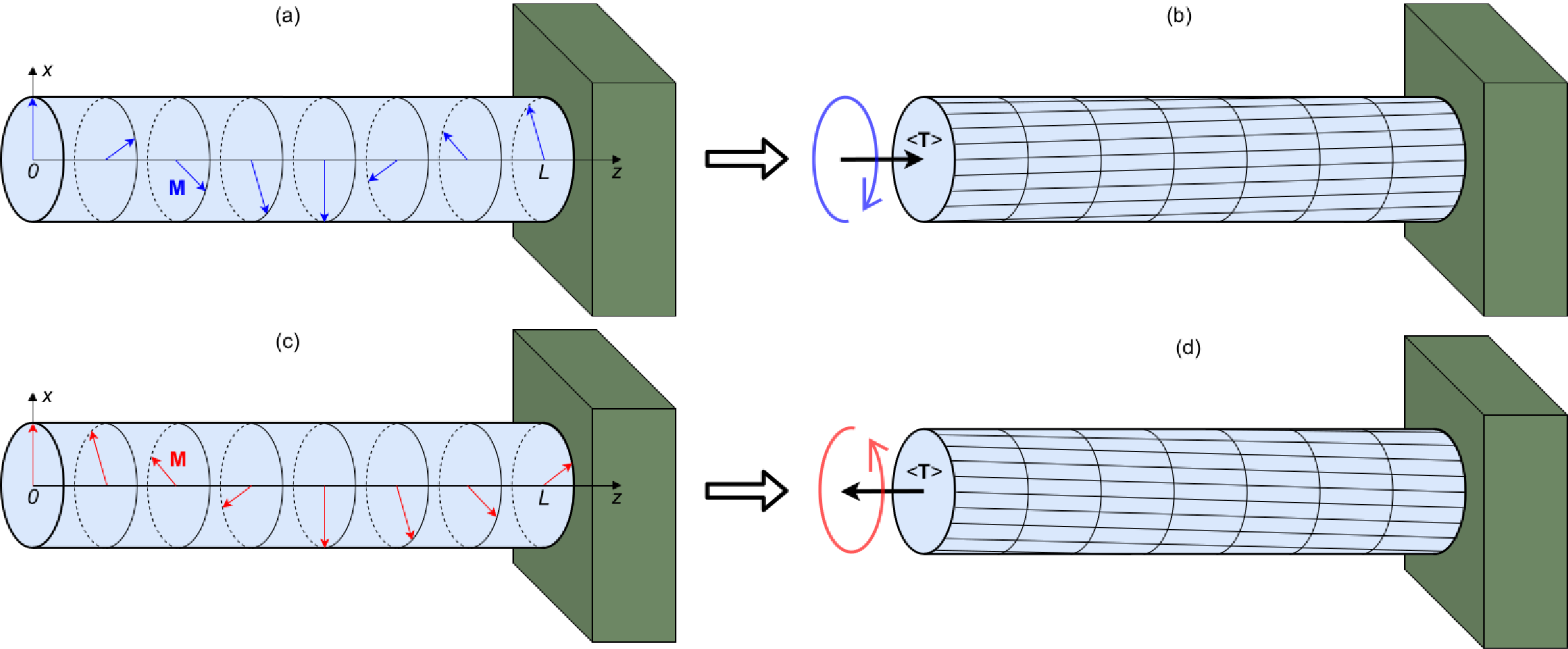}
\caption{\label{fig:1}Schematic illustration of (a,c) helical magnetization distributions $\mathbf{M}$ and (b,d) torsional deformation of a magnetic rod induced by the torque $\langle T_z \rangle$ associated with the corresponding $\mathbf{M}$. Blue arrows correspond to the case $q > 0$ (panel~a) and $\langle T_z \rangle > 0$ (panel~b), while red arrows correspond to $q < 0$ (panel~c) and $\langle T_z \rangle < 0$ (panel~d). In each case, one end of the rod is free ($z = 0$) and the other is rigidly fixed ($z = L$).}
\end{figure*}

The torque $\langle T_z \rangle$ applied at the rod end (Figs.~1b,d) corresponds to the energy~\cite{landau1986}
\begin{equation}
U = \langle T_z \rangle \int\limits_0^L \tau\, dz = \frac{1}{2} \langle T_z \rangle \int\limits_0^L \frac{\partial}{\partial z} (\nabla \times \mathbf{u})_z\, dz, \label{eq:U}
\end{equation}
where $\tau=d\varphi/dz$ and $\varphi$ is the twist angle. Recall~\cite{landau1986} that for elastic torsion of a rod, the displacement field is $\mathbf{u} = (-\tau y (z - L),\, \tau x (z - L),\, 0)$, so that $\nabla \times \mathbf{u} = (-\tau x,\, -\tau y,\, 2\tau (z - L))$. Identifying the energy $U$ with the tiDMI energy,
\begin{equation}
E_{\text{tiDMI}} = S \int\limits_0^L \varepsilon_{\text{tiDMI}}\, dz = S D_1 q \int\limits_0^L \frac{\partial}{\partial z} (\nabla \times \mathbf{u})_z\, dz,
\end{equation}
we obtain the following expression for the constant $D_1$:
\begin{equation}
D_1 = -\frac{\langle Q_{zz} \rangle}{2 V q} = \frac{\hbar^2 N}{32 m} \left( \frac{J}{\varepsilon_F} \right)^2. \label{eq:D1}
\end{equation}
Finally, we determine the twist rate $\tau$. To this end, we add to $U = E_{\text{tiDMI}}$ the elastic energy of the twisted rod,
\begin{equation}
E_{\text{el}} = \frac{1}{2} \int\limits_0^L C \tau^2\, dz,
\label{eq:Eel}
\end{equation}
where $C$ is the torsional rigidity. For a cylindrical rod of radius $R$, $C = \mu \pi R^4 / 2$, with $\mu$ being the shear modulus~\cite{landau1986}. Minimizing the total energy $E = E_{\text{el}} + E_{\text{tiDMI}}$, we find
\begin{equation}
\tau = -\frac{\langle T_z \rangle}{C} = -\frac{\hbar^2 N S q}{16 C m} \left( \frac{J}{\varepsilon_F} \right)^2. \label{eq:tau}
\end{equation}
Note that, as evident from Eq.~\eqref{eq:tau}, the torsion directions of the rod for helices of opposite chirality are opposite, yet the energies of these configurations are identical. Indeed, using Eqs.~\eqref{eq:U}, \eqref{eq:Eel} and~\eqref{eq:tau}, one finds
\begin{equation}
E = -\frac{1}{2} \int\limits_0^L C \tau^2 \, dz,
\end{equation}
so the chiral symmetry of the sample remains unbroken.

Thus, tiDMI arises from the transfer of angular momentum to the lattice upon electron reflection from the magnet boundary, accompanied by a change in electron spin orientation. The constant $D_1$ characterizing the tiDMI energy, as well as the twist rate $\tau$ of the helimagnetic rod, are determined by the square of the ratio of the $s$--$d$ exchange constant to the Fermi energy, $(J / \varepsilon_F)^2$.

\section{Conclusion}

In this work, we have described the microscopic mechanism of torsion-induced Dzyaloshinskii--Moriya interaction and determined the constant $D_1$ entering the general expression for the deformation-induced Dzyaloshinskii--Moriya energy~\eqref{eq:ediDMI}, which is associated with inhomogeneous deformations. We now provide numerical estimates for $D_1$ and the twist rate $\tau$.

Substituting the $s$--$d$ exchange constant and Fermi energy for holmium ($J \approx 0.175\,\text{eV}$~\cite{krizek1975}, $\varepsilon_F \approx 5\,\text{eV}$~\cite{Papaconstantopoulos}) into Eq.~\eqref{eq:D1}, we obtain $D_1 \sim 10^{-9}\,\text{erg/cm}$. Further, assuming a helix period $\lambda\approx 3.5\,\text{nm}$~\cite{koehler1965} and shear modulus $\mu = 2.6 \times 10^{11}\,\text{erg/cm}^3$~\cite{scott1978}, for a rod of radius $R \sim 10\,\text{nm}$ we find $\tau \sim -0.1\,\text{cm}^{-1}$. Thus, the total twist angle $\Delta\varphi = |\tau| L$ of the free end of a rod of length $L = 1\,\mu\text{m}$ does not exceed $10^{-5}$ and is unlikely to be measurable. Moreover, the product $D_1 |\tau| \sim 10^{-10}\,\text{erg/cm}^2$ is 9--10 orders of magnitude smaller than typical Dzyaloshinskii--Moriya energies in ferromagnet/heavy-metal heterostructures~\cite{di2015}. Consequently, tiDMI cannot induce spontaneous torsional deformations and stabilize a helimagnetic order in ferromagnets.

\begin{figure*}[t]
\includegraphics[width = 0.85\textwidth]{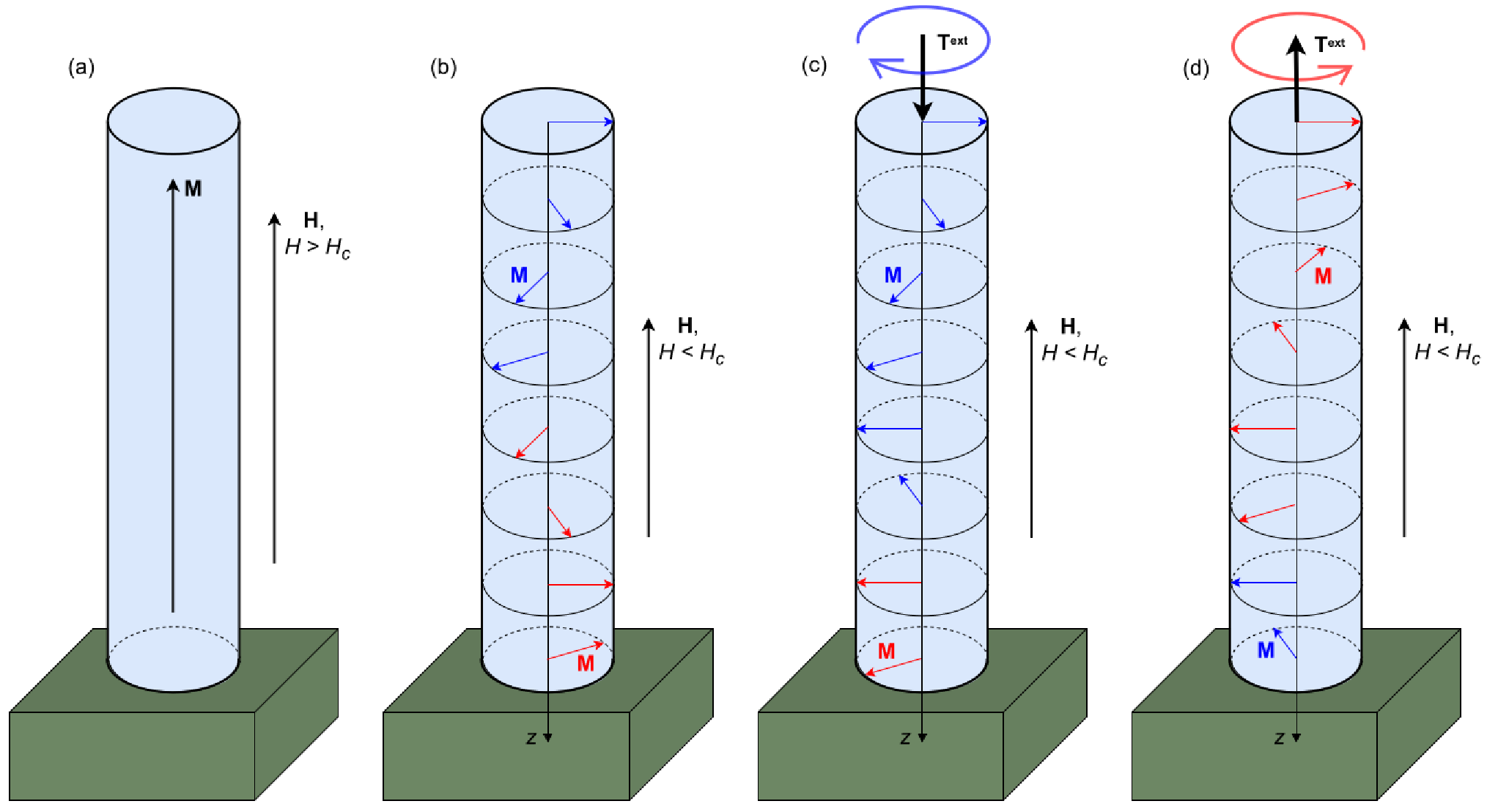}
  \caption{\label{fig:2}%
  Schematic illustration of magnetization distributions $\mathbf{M}$ in a helimagnetic rod maintained at $T < T_N$: (a) uniform $\mathbf{M}$ stabilized when $H > H_c$; (b) right-handed (blue arrows, $q > 0$) and left-handed (red arrows, $q < 0$) magnetic helices forming at $H < H_c$ and occupying equal volumes. Schematic illustration of chiral symmetry breaking induced by an external torque $T_z^{\text{ext}}$ that twists the rod under conditions $T < T_N$ and $H < H_c$: (c) enhanced volume fraction of the right-handed helix for $T_z^{\text{ext}} > 0$; (d) enhanced volume fraction of the left-handed helix for $T_z^{\text{ext}} < 0$. The component of $\mathbf{M}$ parallel to $\mathbf{H}$ in panels b--d is neglected.}
\end{figure*}

However, under certain conditions, tiDMI can lift the chiral degeneracy in helimagnets. Indeed, if external forces are applied to the free end of the rod, inducing torsion, the total energy becomes
\begin{align}
E &= \int\limits_0^L \left( \frac{1}{2} C \tau^2 + (T_z^{\text{ext}} + \langle T_z \rangle) \tau \right) dz \nonumber\\
&= -\frac{1}{2} C L \tau^2, \label{eq:E_total}
\end{align}
where $T_z^{\text{ext}}$ is the external torque, and the twist rate is $\tau = \tau_{\text{in}} + \tau_{\text{ext}}$, with $\tau_{\text{in}} = -\langle T_z \rangle / C$ arising from tiDMI (see Eq.~\eqref{eq:tau}) and $\tau_{\text{ext}} = -T_z^{\text{ext}} / C$ due to external forces. The energy~\eqref{eq:E_total} contains a term proportional to $\tau_{\text{ext}} \tau_{\text{in}}$, whose sign depends on the chirality of the helix. Hence, chiral symmetry is broken, and the energy difference between right-handed ($q > 0$) and left-handed ($q < 0$) helices is
\begin{equation}
\Delta E = E(q) - E(-q) = 4 D_1 V q \tau_{\text{ext}}.
\end{equation}
When the condition
\begin{equation}
4 D_1 S |q| \, \Delta\varphi_{\text{ext}} > k_B T
\end{equation}
is satisfied (where $k_B$ is Boltzmann’s constant, $T$ is temperature, and $\Delta\varphi_{\text{ext}} = |\tau_{\text{ext}}| L$ is the total twist angle induced by external forces), this energy difference should lead to the preferential nucleation of helices of a single chirality. For $\Delta\varphi_{\text{ext}} \sim 1$ and $k_B T \sim 10^{-15}\,\text{erg}$, one requires $S > 10^{-14}\,\text{cm}^2$ (i.e., $R > 1\,\text{nm}$), which is easily achievable.

We therefore propose the following experiment to observe tiDMI. A macroscopic rod is cut from a holmium single crystal, with its axis perpendicular to the close-packed planes and aligned with the rod axis. At low temperature, an external magnetic field is applied along the rod axis. The temperature $T$ should be chosen below the Néel temperature $T_N = 132\,\text{K}$~\cite{koehler1965} at which helical magnetic order forms in holmium. Initially, the magnetic field $H$ exceeds the critical value $H_c$ required to stabilize a uniform magnetization along the field direction (Fig.~2a). Upon reducing the field below $H_c$, helical magnetization distributions develop in the rod. Due to the intrinsic chiral symmetry of holmium, the volumes occupied by left- and right-handed helices are equal~\cite{cowley1988} (Fig.~2b). However, if the sample is subjected to torsional deformation induced by external forces, then upon reducing the magnetic field below $H_c$, helices with predominantly one sign of $q$ will nucleate, and the sample will become magnetochiral (Fig.~2c). Moreover, the sign of the magnetic chirality is determined by the direction of torsion (cf. Figs.~2c and~2d). This transformation can be detected via thermal neutron diffraction~\cite{Maleev2002}.  Since the critical field is large enough ($H_c\approx120$~kOe at $T=4$~K~\cite{Jensen1991}), a similar experiment can be performed by varying the temperature near $T_N$: cooling the sample below $T_N$ under torsional strain will lead to the formation of a  magnetochiral state.
\vspace{2pt}
\begin{acknowledgments}
\vspace{2pt}
The work was supported by the Russian Science Foundation (Grant No.~25-22-00126). We acknowledge very useful discussions with E.~A.~Karashtin.
\end{acknowledgments}

\bibliography{apssamp}

\end{document}